# Band gaps atlas for photonic crystals having the symmetry of the kagomé and pyrochlore lattices

**Angel J. Garcia-Adeva**

*Departamento de Física Aplicada I, E. T. S. de Ingeniería de Bilbao, UPV/EHU, Alda. Urquijo s/n, 48013 Bilbao (Spain)*
*wuzgaada@lg.ehu.es*

**Abstract:** A new family of two- and three-dimensional photonic crystals with the symmetry of the kagomé and pyrochlore lattices that exhibit interesting photonic band structures are reported. From large complete photonic band gaps in three dimensions to enormous partial gaps in two dimensions for certain polarizations occurring at feature sizes that make these lattice amenable of fabrication, the results described below sure make this new family of photonic crystals very promising for potential applications and very interesting from the fundamental point of view.

**Introduction**

Photonic crystals are engineered periodic structures made of two or more materials with very different dielectric constants. They have generated an ever-increasing interest in the last twenty years because of their potential to control the propagation of light to an unprecedented level [1–17]. When an electromagnetic wave (EM) propagates in such a structure whose period is comparable to the wavelength of the wave, interesting phenomena occur. Among the most interesting ones are the possibility of forming a complete photonic band gap (CPBG), that is, a frequency range for which no photons having frequencies within that range can propagate through the photonic crystal, to localize light by introducing several types of defects in the lattice, or enhancing certain non-linear phenomena due to small group velocity effects.

The problems one faces in order to implement these advantageous properties into technological applications are mainly two: on the one hand, the scarcity of three-dimensional symmetries with a CPBG that are known. Indeed, most known structures with a CPBG are based on the face-centered cubic (fcc) or diamond lattices and their derivatives [18], such as the A7 structure or graphite [2,19,20]. On the other hand, manufacturing these lattices requires working on sub-micron lengthscales so ingenious innovations were required in order to actually fabricate them [21–25]. Also, this has caused a lot of effort being devoted to investigate photonic-crystal slabs, i.e., dielectric structures that are periodic in two dimensions and uses index guiding to confine light in the third dimension [12,15,16]. Therefore, identifying new two- and three-dimensional photonic crystal structures is an important task even at the technological level.

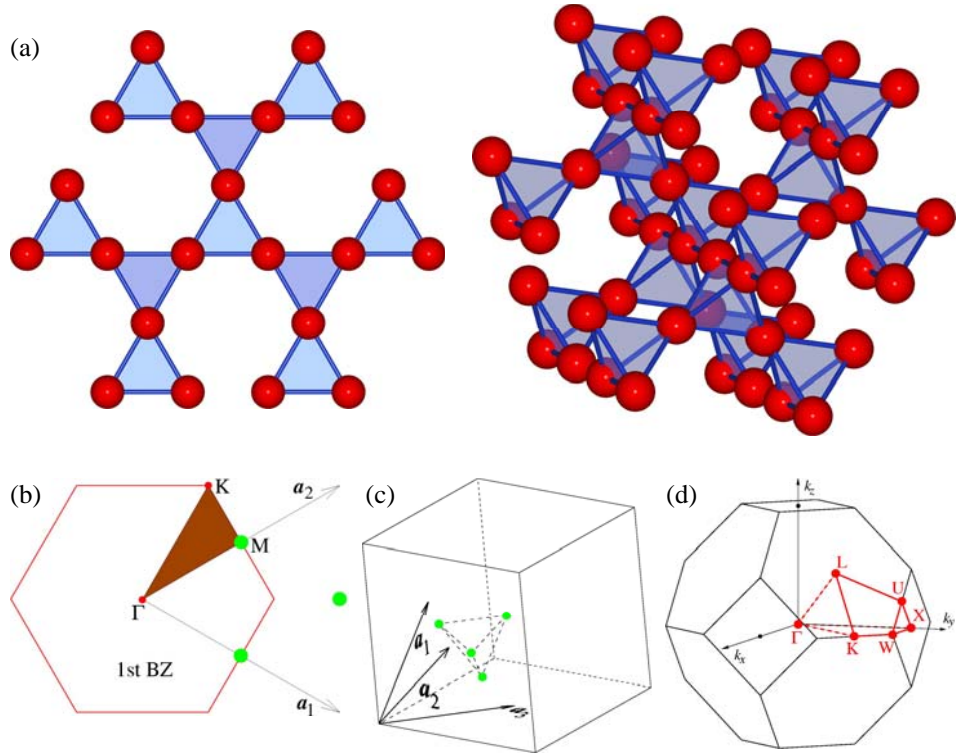

Fig. 1. A primer on the kagomé and pyrochlore lattices. (a) Pictorial representation of the kagomé (left) and pyrochlore (right) lattices. (b) Basis vectors ($a_1$ and $a_2$), atomic positions (green dots), and 1$^{st}$ Brillouin zone (red hexagon) of the kagomé unit cell. The atomic positions are given by (0,0), (1/2,0), and (0,1/2) in the ($a_1$, $a_2$) basis. The special symmetry points Γ, M, and K delimit the irreducible part of the 1$^{st}$ BZ (shaded in brown). Because of symmetry arguments, only *k*-vectors in that triangle are necessary for the band calculations.(c) Basis vectors ($a_1$, $a_2$, and $a_3$) and atomic positions (green spheres) of the pyrochlore unit cell. The atomic positions in this basis are given by (1/2, 1/2, 1/2), (0, 1/2, 1/2), (1/2, 0, 1/2), and (1/2, 1/2, 0). (d) 1$^{st}$ BZ of the pyrochlore lattice. The red polyhedron defined by the special symmetry points Γ, L, K, W, X, and U delimits the irreducible part of the 1$^{st}$ BZ. It is important to notice that neither the kagomé nor the pyrochlore lattices are Bravais lattices, as it is not possible to find a unit cell that contains a single atom.

The main purpose of this paper is to explore the photonic properties of various topologies based on the kagomé and pyrochlore symmetries. The kagomé and pyrochlore lattices (see Fig. 1a) are well known to the magnetic properties practitioners because of their intriguing properties that include geometrical frustration and anomalous electronic behaviors [26–33]. The two-dimensional kagomé lattice is a lattice of corner-sharing triangles with "atoms" located at the corners of the triangles. It is important to stress at this point that the photonic properties of certain topologies with the kagomé symmetry (mainly a kagomé lattice of dielectric cylinders in air) have been studied recently by other authors [34–38]. However, the present study covers various aspects not discussed in those works (as they are the band gap maps for several values of the dielectric constant, the photonic density of states, etc.). The three-dimensional pyrochlore lattice is a lattice of corner sharing tetrahedral with "atoms" located at the corners of the tetrahedral. Two important characteristics of these lattices are that they are based on triangular elementary units and that they are very sparsely connected, which are known to favor the formation of photonic band gaps [1]. The study reported below shows that photonic crystals based on these symmetries are very promising candidates for photonic applications and quite interesting from the fundamental point of view.

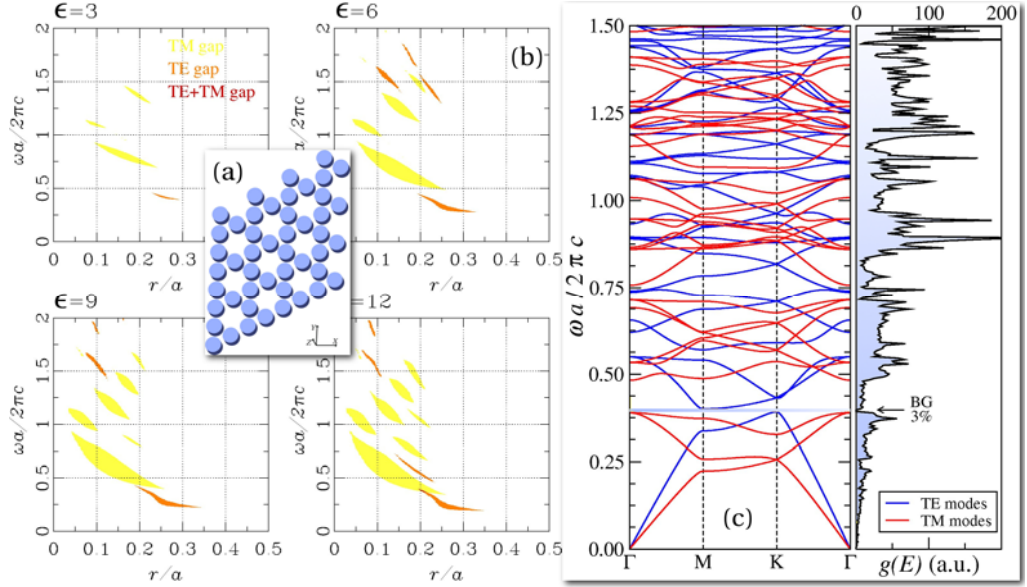

Fig. 2. Band gap map at selected values of ε for the kagomé lattice of dielectric rods in air. (a) Pictorial representation of the lattice at $r/a = 0.19$ (radius of the dielectric cylinders at which the maximum BG is attained for ε = 12). (b) Band gap map for this structure. (c) Band structure and density of states at $r/a = 0.19$ for ε = 12. The blue stripe delimits the photonic band gap. 32 TM bands and 32 TE bands were used for all the band calculations. A 32×32 grid was used for discretizing the lattice. 200 $k$-vectors along the contour of the irreducible part of the 1$^{st}$ BZ (see Fig. 1) were used to compute the bands and 3000 $k$-vectors in the 1$^{st}$ BZ were randomly selected to calculate the density of photonic states.

**Details of the calculations and results**

For the present work, we considered perfectly periodic infinite lattices and the band structure of the electromagnetic field propagating in these structures was calculated. To this end, the eigenmodes of Maxwell's equations with periodic boundary conditions were computed by preconditioned conjugate-gradient minimization of the block Rayleigh quotient in a planewave basis, using the freely available software package MIT Photonic-Bands (MPB) [39]. Five different topologies with the symmetry of the kagomé or pyrochlore lattices were studied:

- In two dimensions, a kagomé lattice of dielectric rods in air (Fig. 2a) and a kagomé lattice of air rods in dielectric (Fig. 3a) were considered. The dielectric material was assumed to be isotropic and non-magnetic. The band structures for these lattices were calculated for different values of the dielectric constant ranging from ε = 3 to ε = 12. Also, the band structure was calculated for several values of the ratio $r/a$, where $r$ is the radius of the rod and $a$ is the lattice parameter (see Fig. 1b), ranging from 0.01 to 0.5, so that the band gap map for a given ε could be constructed [1]. For these two-dimensional calculations, the EM wave was assumed to propagate in the plane of the lattice. Details about the computations can be found in the caption of Fig. 2.

- In three dimensions, three different possibilities were studied. The first one corresponds to a pyrochlore lattice of dielectric spheres in air (see Fig. 4a). The second one corresponds to the inverse structure of air spheres embedded in a dielectric material (see Fig. 5a). Air cylinders along the lines that join the atomic positions in the pyrochlore lattice form the third one (see Fig. 6a). This later one is specially interesting from the manufacturing point of view as it can be made by directly drilling the air veins in a dielectric material, in a similar fashion to the well known Yablonovite structure [9], as will be explained below. The same computational scheme as for the two-dimensional lattices was applied to the three-dimensional ones (see Figs. 1c, 1d, and the caption of Fig. 4) with two differences:

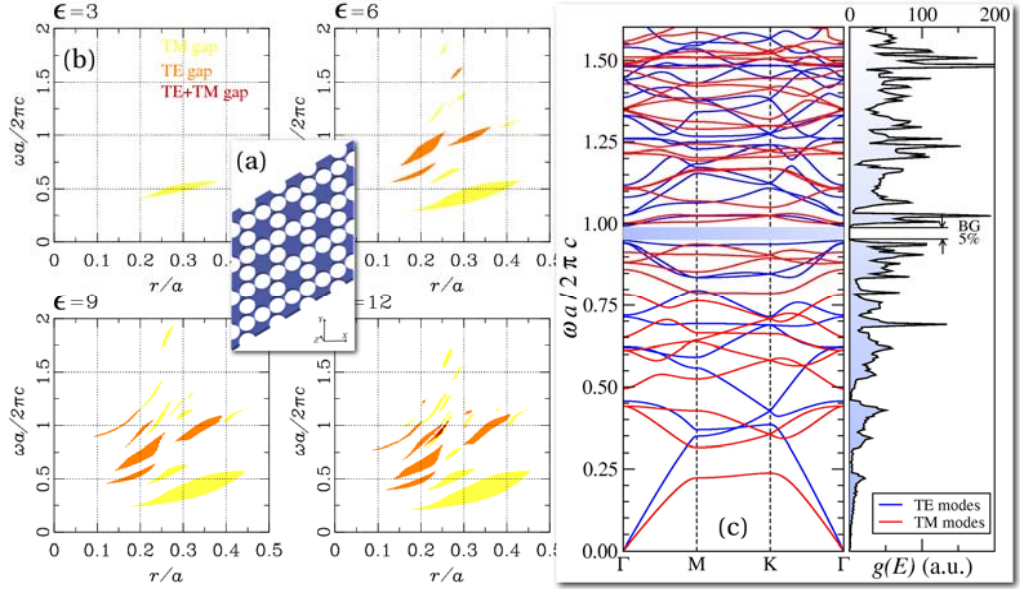

Fig. 3. Band gap map at selected values of ε for the kagomé lattice of air rods in dielectric. (a) Pictorial representation of the lattice at $r/a = 0.24$ (radius of the air cylinders at which the maximum BG is attained for ε = 12). (b) Band gap map for this structure. (c) Band structure and density of states at $r/a = 0.24$ for ε = 12; the blue stripe delimits the photonic band gap. See the caption of Fig. 2 for the computational details.

arbitrary propagation directions of the EM wave were considered and the dielectric constant range spans only from 6 to 12, due to the longer CPU times required for three-dimensional calculations.

It is worthy to say some words about the various quantities reported below. First of all, as it is customary, results are quoted in terms of the dimensionless frequency $\omega a/2\pi c$, with ω the wave number of the EM field, $a$ the lattice parameter, and $c$ the speed of light in vacuum. With regards to the band gap maps presented below, it should be stressed that this is an important quantity as it readily provides a visual feedback of the frequency ranges in which photon propagation is forbidden for given values of the dielectric constant and the size of the feature defining the lattice. For a given topology, the band gap map is simply obtained by plotting the locations of the photonic gaps found in the band structure (if any) as a function of one or more parameters (ε and $r/a$ in our case). One obvious advantage of this type of visualization is that it is very easy to identify the optimal geometrical parameters of the lattice in order to maximize the width of the band gap. Also, if fabrication at the optimal value of $r/a$ is not feasible for any reason, we can readily know from this map whether fabrication at a larger feature size will still give rise to a CPBG, even at the cost of a smaller gap. Regarding the size of the gap, it should be noted that it is expressed in terms of the mid-gap to band gap ratio, $\Delta\omega/\omega_m$, which is a quantity that does not change with the frequency of the EM field. Another interesting quantity we can readily compute from the band structure is the density of photonic states. Some examples of this quantity are also given in the figures below. This quantity is obtained by randomly selecting a number of k-vectors in the 1$^{st}$ BZ and calculating the corresponding eigenfrequencies. The density of states is then obtained by counting the number of modes with frequencies in the range from ω to $\omega+\Delta\omega$, where $\Delta\omega$ is the frequency spacing (typically 0.001 in our calculations).

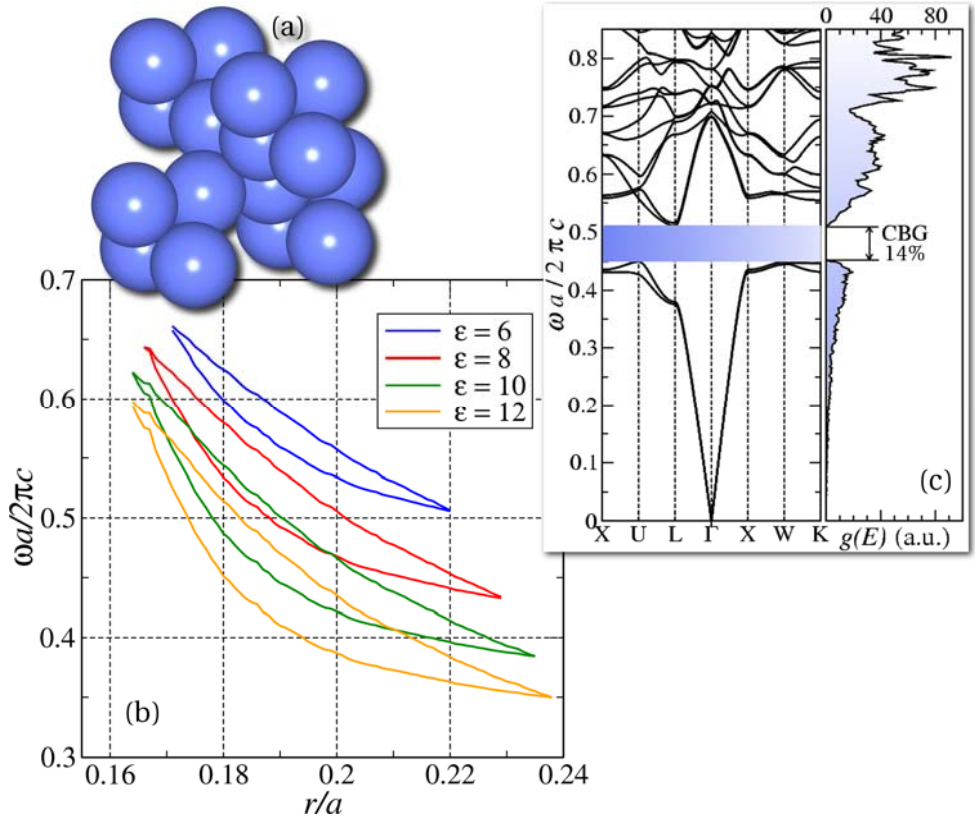

**Fig. 4. Band gap map at selected values of ε** for the pyrochlore lattice of dielectric spheres in air. (a) Pictorial representation of the lattice at $r/a = 0.19$ (radius of the dielectric spheres at which the maximum PBG is attained for ε = 12). (b) Band gap map for this structure. (c) Band structure and density of states at $r/a = 0.19$ for ε = 12. 16 bands were computed for all the calculations. A 24×24×24 grid was used for discretizing the lattice. 500 **k**-vectors along the contour of the irreducible part of the 1st BZ were used to compute the bands and 5000 **k**-vectors in the 1st BZ were randomly selected to calculate the density of photonic states.

Let us now review the main results for the topologies studied in this work, starting with the two-dimensional lattices. A summary of the most interesting results and the values of some parameters important for fabrication can be found in Table 1. Fig. 2b shows the band gap map of the kagomé lattice of dielectric rods in air for selected values of ε. As can be seen from that figure, this geometry displays a wealth of PBG for TM polarization (even for values of ε as small as 3) and some smaller PBG for TE polarization. Moreover, there is a small overlap between TM and TE gaps that gives rise to a PBG for all polarizations [40] with a maximum at $r/a = 0.21$ for ε = 12 (dielectric rods made of Si). However, this gap is small ($\Delta\omega/\omega_m \sim 3\%$) and extends over a narrow range of the dielectric rod radii so we cannot expect it to be very robust upon disorder always present in real systems. On the other hand, the gaps for TM polarization are large and well separated from the TE ones, which could make this structure very useful for polarization dependent applications. In particular the TM gap between TM bands 3 and 4 is huge (539 nm at λ = 1.3 μm for ε = 12) and extends over a large $r/a$ interval. As an illustrative example, we have depicted in Fig. 2c the band structure of this lattice for ε = 12 at the value of $r/a$ that maximizes the PBG for all polarizations and the corresponding density of states. The band structure for this lattice shows some interesting features. For example, the extrema of the band are not always located at the zone boundaries. This is a direct consequence of the non-Bravais character of this lattice, i.e., having more than one atom per unit cell. Also, some of the bands are very flat in some directions (noticeably bands 3 and 9 for both TM and TE polarizations in the ΓM direction and the 2nd TM band in the MK direction for small values of ω and especially bands with a large band index), which will give rise to the small group velocity effect associated with the enhancement of certain processes such as the stimulated emission, sum-frequency generation, etc, as discussed in [14]. The density of photonic states is also shown for comparison. At low frequencies, this quantity is linear with a slope that depends on ε, due to the fact that long-wavelength modes "are insensitive" to the inhomogeneities of the structure. As the frequency is increased, this quantity follows a more complicated behaviour due to the multiple scattering by the inhomogeneous structure. It is also apparent the small PBG for all polarizations mentioned above. Also, the small velocity effects mentioned before clearly show up in this quantity as very narrow peaks occurring most noticeably for large values of ω.

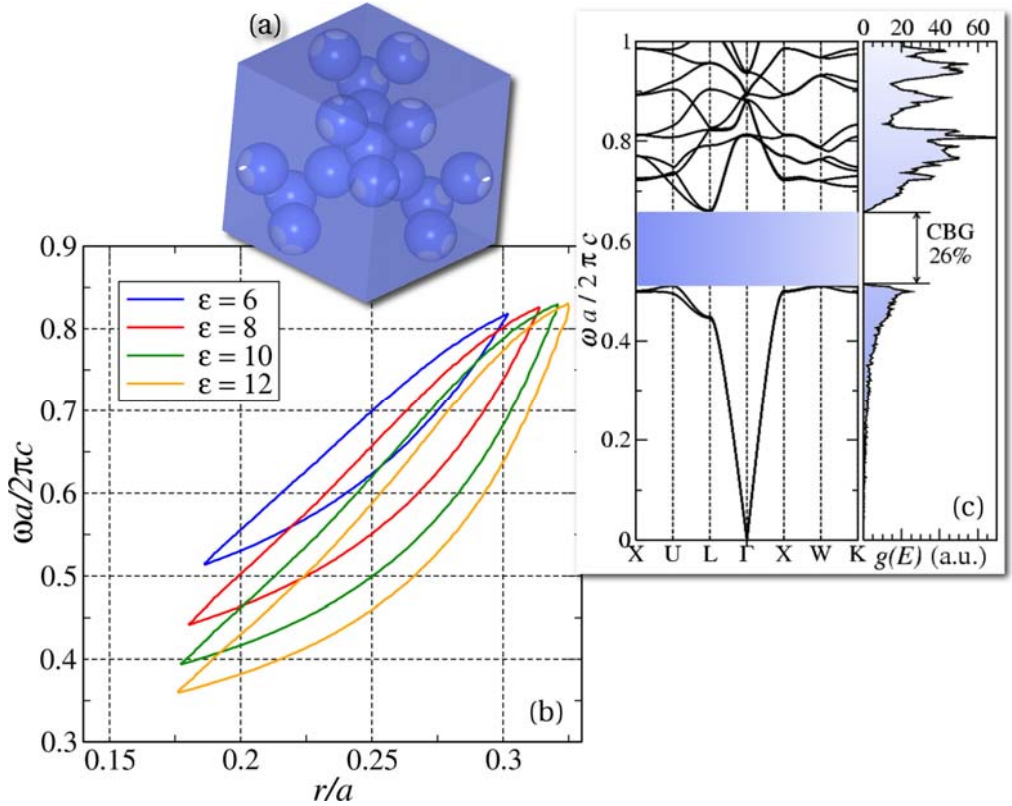

Fig. 5. Band gap map at selected values of ε for the pyrochlore lattice of air spheres in dielectric. (a) Pictorial representation of the lattice at $r/a = 0.27$ (radius of the air sphere at which the maximum PBG is attained for ε = 12). (b) Band gap map for this structure. (c) Band structure and density of states at $r/a = 0.27$ for ε = 12. See the caption of Fig. 4 for the computational details.

The band gap map for the kagomé lattice of air rods in dielectric shows even more interesting features than the previous geometry does. Results for this lattice are shown in Fig. 3b and some data are also summarized in Table 1. There are again a large number of TM gaps. In particular, there is an enormous TM gap between the 1$^{st}$ and 2$^{nd}$ TM bands (668 nm at λ = 1.3 μm for ε = 12) that extends over a very wide $r/a$ range. However, in contrast with the previous geometry, now there are also various large TE gaps. For example, the TE gap occurring between the 6$^{th}$ and 7$^{th}$ TE bands has a $\Delta\omega/\omega_m = 18\%$ (240 nm at λ = 1.3 μm for ε = 12). What is even more interesting is the appearance of two PBG for all polarizations at large frequency values. The lower one, between bands 21 and 22, has a $\Delta\omega/\omega_m = 5\%$. The important point about this gap is that due to the large frequency value at which it appears ($\omega_m a/2\pi c \sim 1$), the lattice spacing required for applications at 1.3μm would be $a \sim 1.3$ μm and the diameter of the rods $r \sim 0.6$ μm, even larger than those for the honeycomb lattice [1]. The band structure for this lattice at the value of $r/a$ that maximizes the complete gap (Fig. 3c) shows even more interesting features than the previous geometry. Indeed, the group velocity anomaly is even more noticeable for various TE and TM bands (most noticeably TM bands 1, 6, 9, and 11, TE bands 2, 5, 9, and 10 in the MK direction, and other higher energy bands for both polarizations). Again, this effect is clearly seen in the density of states. This time, however, it occurs very markedly around the PG.

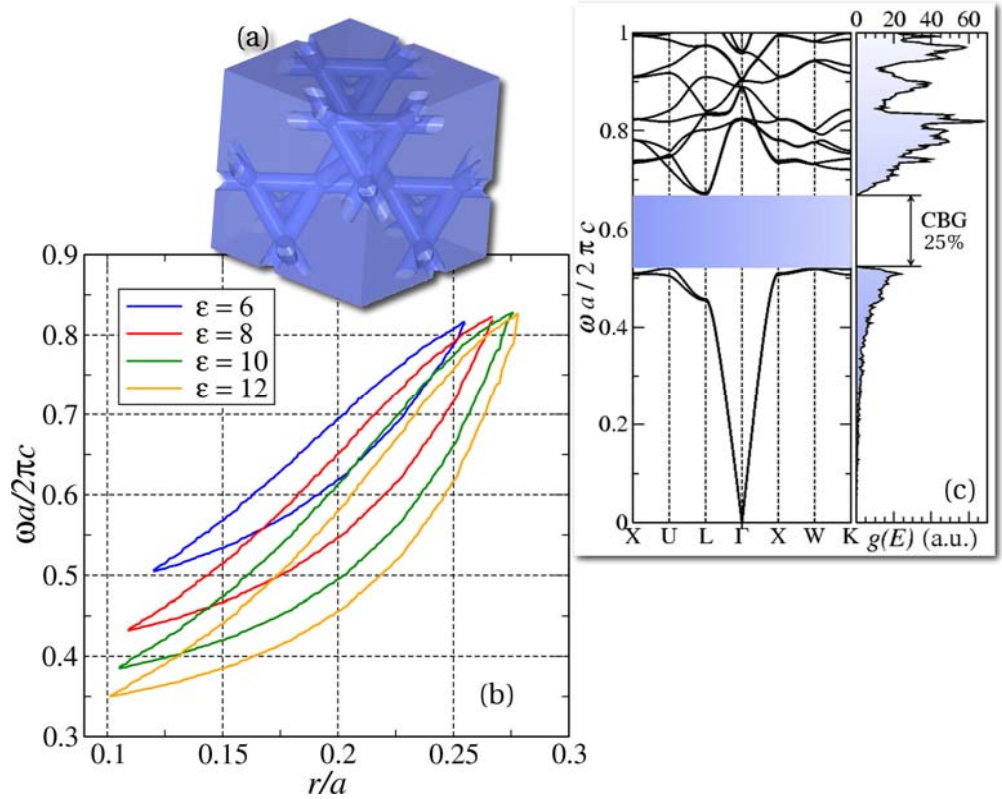

Fig. 6. Band gap map at selected values of ε for the pyrochlore lattice of air rods in dielectric. (a) Pictorial representation of the lattice at $r/a = 0.10$. (b) Band gap map for this structure. (c) Band structure and density of states at $r/a = 0.24$ for ε = 12; the gray stripes delimit the photonic band gap. See the caption of Fig. 4 for the computational details.

Let us now turn our attention onto the results for the 3D pyrochlore lattices. The band gap maps in this case are much simpler than for the 2D lattices but, at the same time, more appealing from the applications point of view. The most interesting result is that the three topologies studied in this work exhibit a complete PBG between bands 2 and 3 that extends over a relatively wide range of $r/a$ values. The lattice of dielectric spheres in air (see Fig. 4b) exhibits a nice size gap with $\Delta\omega/\omega_m \sim 14\%$ for ε = 12 (179 nm gap at λ = 1.3 μm) at $r/a = 0.19$. However, fabricating the corresponding photonic crystal would require making spheres of ~ 0.2 μm diameter, which still constitutes a formidable task. On the other hand, the inverse lattice made of air spheres in dielectric (see Fig. 5b) exhibits a large gap comparable to that of the inverse diamond lattice (which to our knowledge is the lattice displaying the biggest photonic gap) with a band gap to mid-gap ratio ~ 26% for ε = 12 (344 nm gap at λ = 1.3 μm) at $r/a = 0.27$. In this latter case, the lattice spacing required for applications at 1.3 μm is $a \sim 0.76$ μm and the diameter of the air spheres is ~ 0.41 μm. These figures are more favourable from the manufacturing point of view than for the direct lattice. The band gap map for the pyrochlore lattice of air rods in dielectric (see Fig. 6b) is quite similar to the previous one, though shifted towards smaller values of $r/a$. The maximum bandgap to midgap ratio in this case is 25% for ε =

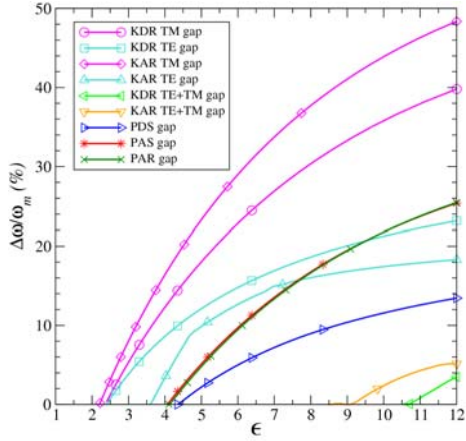 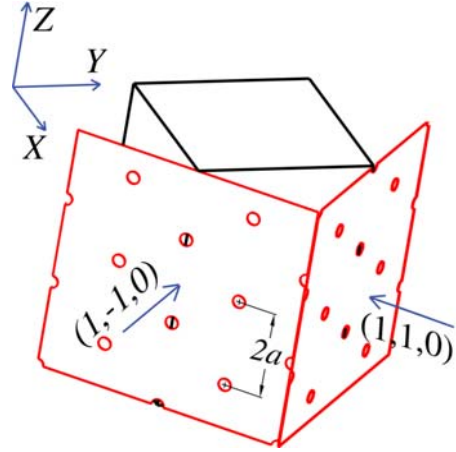

Fig. 7. Evolution of the bandgap-to-midgap ratio with ε for the structures studied in this work. See the legend of Table 1 for the notations used in the legend of this figure.

Fig. 8. Possible fabrication pathway of the pyrochlore lattice of air rods studied in this work. See text for details.

12 (336 nm gap at $\lambda = 1.3$ μm) at $r/a = 0.23$. This implies drilling channels of diameter 0.36 μm for applications at $\lambda = 1.3$ μm.

If we now take a look to the band structure for these lattices ((c) panel in Figs. 4–6), we can notice that they are not that different from each other (and, in turn, they are not that different from the one for the diamond lattice either), especially for the lower bands. Interestingly, bands 2, 3, and 4 are very flat in the XW and WK directions for the three lattices, which surely gives rise to interesting small group velocity phenomena as mentioned for the 2D lattices. The density of photonic states, on the other hand, provides again information in a very intuitive fashion about the propagation of an EM field through these lattices: at low frequencies, we can observe the quadratic behaviour of this quantity, exactly as for EM waves propagating in a 3D homogeneous material, for the reasons explained above. At higher frequencies, the periodic character of the medium clearly shows up as non-trivial features of this quantity, including some peaks associated to flat bands and small group velocity, as explained before.

We have also performed a study of the evolution of the band gap size with varying ε at the value of $r/a$ that maximizes the corresponding gap for ε = 12 and the results are displayed in Fig. 7. For the 2D lattices, we only studied the evolution of the largest gap occurring for each polarization and that of the total gap (see Table 1). As briefly mentioned before, the total gaps for the 2D kagomé lattices are very small and only exist above very large values of ε (ε ~10.5 for the kagomé lattice of dielectric rods and ε ~ 9 for the kagomé lattice of air rods in dielectric). Therefore, these gaps are not very robust and it is very likely that, in real systems, they do not show up because of the disorder always present in real structures that tends to destroy the gap. The partial gaps, however, exist down to ε ~ 2–2.5, except for the first TE gap in the kagomé lattice of air rods in dielectric, which shows a crossover to a linear e dependence for ε < 4.5 and disappears for ε ~ 3.5. Even though this is an interesting behavior that deserves further study, we do not have an explanation for it at this point. With regards to the gaps for the 3D lattices, the one for the pyrochlore lattice of dielectric spheres subsists down to ε ~ 4.4, whereas the evolution of the gap size with ε for both the pyrochlore lattices of air spheres and air rods in dielectric survive down to ε ~ 4 and, actually, the curves that describe their behaviors are almost identical.

Even though this is not the main topic of the present manuscript, I would not like to finish without making some comments about the plausibility of fabricating the structures studied in this work. A more detailed discussion about this subject will be given elsewhere [41]. Regarding the 2D lattices, I think the feature size and lattice spacing required for applications at $\lambda = 1.3$ μm (see Table 1) are well in the reach of current technological capabilities. In some cases, these sizes are even more favourable than the corresponding ones for photonic crystals based on the triangular lattice, which are routinely made in the laboratory, especially for the kagomé lattice of air rods in dielectric. Therefore, it should not be too difficult to fabricate this structure by using currently avalaible lithographic techniques [42]. Fabrication of the 3D structures, however, it is very likely more complicated. Fortunately, there are some geometrical features of the pyrochlore lattice that could simplify this task. In particular, the pyrochlore lattice can be built by stacking alternating triangular and kagomé lattices in an ABAB… sequence [41]. Given the fact just mentioned that triangular layers are relatively easy to make and that kagomé layers have been made by sedimentation of a binary mixture of colloidal particles [43], it is likely that photonic crystals with topologies based on spheres and the symmetry of the pyrochlore lattice could be fabricated by using sedimentation of colloidal particles in a layer-by-layer fashion. On the other hand, the pyrochlore lattice of air rods is amenable of fabrication for applications in the microwave regime by a technique analogous to the one that allowed to make the well known yablonovite structure some years ago [9], namely, by drilling or etching the air channels in a block of dielectric. In the visible range, ion beams or holographic techniques could probably be used to drill the channels. One possible pathway that can be followed to create the channel structure of the right geometry is depicted in Fig. 8. It consists on using two masks containing a triangular pattern which are arranged perpendicular to each other as shown in the figure and with the common edge parallel to one of the edges of the dielectric block (the $X$ direction). The channels are then drilled along perpendicularly to the masks. Later, the masks are oriented with the common edge parallel to the $Y$ and $Z$ directions and the drilling operation is repeated. Of course, the procedure is not as simple as for the yablonovite structure, but the size of the gap that can be obtained in this way makes it worth trying.

**Conclusions**

In this work, I report on a new family of geometries based on the kagomé and pyrochlore lattices that exhibit important photonic properties that make them relevant from the applications point of view. Two structures have been investigated in two dimensions, a kagomé lattice of dielectric rods in air and the inverse lattice consisting on a kagomé lattice of air rods in a dielectric matrix. The band gap maps for different values of the $r/a$ ratio and index contrast for EM waves propagating in the plane of the lattice are very rich and exhibit numerous band gaps for TM and TE polarizations. In the direct lattice, these gaps do not overlap, which makes this structure a good candidate for polarization dependent devices. However, the inverse lattice of air rods in dielectric exhibits a complete, albeit small, photonic gap at the dimensionless frequency $\omega a / 2\pi c \approx 1$, which implies an advantageous $a \approx \lambda$ lattice parameter for actual device fabrication. In three dimensions, three different possibilities have been studied, a pyrochlore lattice of dielectric spheres in air, the inverse structure consisting on a pyrochlore lattice of air spheres in a dielectric matrix, and the structure formed by air cylinders that connect the centers of the "atomic positions" in the pyrochlore lattice embedded in a dielectric medium. These three cases exhibit large complete band gaps in a wide range of the $r/a$ ratio and index contrast. Especially interesting is the case of air spheres in dielectric, which exhibits a complete band gap with a maximum band gap to midgap ratio of 26% for $\varepsilon = 12$. The lattice made of air cylinders in dielectric, on the other hand, is very convenient from a fabrication point of view and also exhibits a complete band gap of the same size as the the previous topology. The lattice parameter and feature size required to fabricate devices based on these symmetries are also very convenient, especially for the pyrochlore lattice of air spheres in dielectric and the tubular structure of air rods in dielectric.

Of course, this is not a closed work. There is a wealth of additional studied that could be performed in order to further assess the usefulness of these new symmetries for applications. For example, one could try to optimize the size of the gaps by using coated spheres or cylinders or using slightly different topologies. Additionally, it would be very interesting to study how robust the photonic gaps are against positional and/or compositional disorder that will always occur in real systems. Also, it is crucial to consider how well these geometries could confine and guide light by introducing different kinds of organized defects such as channels or point defects.

In conclusion, I hope that this work serves the purpose of stimulating other researchers in this field to further explore the properties of materials based on the kagomé and pyrochlore lattices for prospective photonic applications.

### Acknowledgments

I would like to thank Dave L. Huber, Joaquín Fernández, and Rolindes Balda for useful comments about this manuscript. This work has been supported by the Spanish MEC under the "Ramón y Cajal" programme.

Table 1. Various important physical parameters of the photonic crystals studied in this work for ε = 12.

| Lattice type | Polarization | LB | UB | $\Delta\omega/\omega_m$ (%) | $\omega_m a/2\pi c$ | $r/a$ (max) | $r_{min}$ | $r_{max}$ | $\Delta\omega_{max}$ (1.3μm) | $a$ (1.3μm) | $r$ (1.3μm) |
|---|---|---|---|---|---|---|---|---|---|---|---|
| KDR | TE | 1 | 2 | 23 | 0.25 | 0.25 | 0.19 | 0.35 | 305 | 0.32 | 0.08 |
|  | TE | 3 | 4 | 5 | 0.56 | 0.20 | 0.14 | 0.23 | 63 | 0.72 | 0.14 |
|  | TM | 3 | 4 | 40 | 0.58 | 0.12 | 0.03 | 0.26 | 539 | 0.76 | 0.09 |
|  | TM | 10 | 11 | 12 | 1.13 | 0.12 | 0.08 | 0.16 | 156 | 1.47 | 0.18 |
|  | TM | 19 | 20 | 9 | 1.33 | 0.15 | 0.13 | 0.18 | 120 | 1.72 | 0.27 |
|  | TE+TM | 4 | 5 | 3 | 0.38 | 0.21 | 0.19 | 0.22 | 42 | 0.49 | 0.10 |
| KAR | TE | 6 | 7 | 18 | 0.69 | 0.23 | 0.13 | 0.25 | 240 | 0.89 | 0.20 |
|  | TE | 9 | 10 | 5 | 0.95 | 0.25 | 0.21 | 0.27 | 71 | 1.22 | 0.30 |
|  | TE | 10 | 11 | 8 | 0.85 | 0.20 | 0.16 | 0.25 | 97 | 1.10 | 0.22 |
|  | TM | 1 | 2 | 48 | 0.40 | 0.34 | 0.17 | 0.46 | 668 | 0.52 | 0.18 |
|  | TM | 4 | 5 | 6 | 1.04 | 0.41 | 0.21 | 0.33 | 72 | 1.35 | 0.55 |
|  | TM | 12 | 13 | 7 | 0.95 | 0.25 | 0.21 | 0.27 | 89 | 1.23 | 0.31 |
|  | TE+TM | 21 | 22 | 5 | 0.94 | 0.25 | 0.23 | 0.27 | 67 | 1.23 | 0.30 |
|  | TE+TM | 24 | 25 | 2 | 0.98 | 0.24 | 0.24 | 0.25 | 23 | 1.28 | 0.31 |
| PDS | n.a. | 2 | 3 | 14 | 0.45 | 0.19 | 0.16 | 0.24 | 179 | 0.59 | 0.11 |
| PAS | n.a. | 2 | 3 | 26 | 0.59 | 0.27 | 0.18 | 0.32 | 344 | 0.76 | 0.21 |
| PAR | n.a. | 2 | 3 | 25 | 0.60 | 0.23 | 0.10 | 0.28 | 336 | 0.78 | 0.18 |

Legend: KDR, kagomé lattice of dielectric rods; KAR, kagomé lattice of air rods; PDS, pyrochlore lattice of dielectric spheres; PAS, pyrochlore lattice of air spheres; PAR, pyrochlore lattice of air rods; LB, lower band defining the PBG; UB, upper band defining the PBG; $r/a$ (max), value of $r/a$ at which the maximum midgap to band gap ratio occurs; $r_{min}$ and $r_{max}$ are the minimum and maximum values of $r/a$ between which there is a band gap, respectively; $\Delta\omega_{max}$ (1.3μm) stands for the maximum value of the band gap (in nm) at λ = 1.3 μm; $a$ and $r$ are the lattice parameter and radius of the feature (sphere, rod,…) in μm at which that maximum occurs for λ = 1.3 μm, respectively.